# Visualization of Suppressed Shock Wave/Turbulent Boundary Layer Interaction Using Cryogenic Wall Cooling

Yuma Miki (三木佑真),[1, a)] Leo Ando (安藤嶺央),[1] Maria Acuña[2] and Kiyoshi Kinefuchi (杵淵紀世志)[1]

1 Department of Aerospace Engineering, Nagoya University, Nagoya, Aichi 464-8603, JAPAN

2 Department of Mechanical and Aerospace Engineering, New York University Tandon School of Engineering, Brooklyn, NY 11201, USA

a) Author to whom correspondence should be addressed: miki.yuma.n4@s.mail.nagoya-u.ac.jp

**Abstract**

To investigate the wall-cooling effects on shock wave/turbulent boundary layer interaction (SWTBLI) with limited experimental data, a supersonic wind tunnel wall was cooled using liquid nitrogen as the cryogenic coolant. Under the condition of the mainstream Mach number, in the range of 2.02–2.04, the wall temperature was cooled to 88−92 K, corresponding to a wall-to-recovery temperature ratio ($T_w/T_r$) of 0.31–0.33. The flow structures with and without wall cooling were observed using the schlieren method. The reflected shock motion or interaction length $L_{int}$ in the schlieren image suggested that boundary layer separation was suppressed under the cooling condition in relation to that under the non-cooling condition, and the suppressed ratio of the cooling-to-uncooling interaction length was approximately 0.60–0.72. Additionally, while the mainstream state and wall temperature near the separation were constant, a gradual change in the separated flow field was observed under the wall-cooling condition. This was due to the slow wall temperature increase in the upstream wall of the separation region, where the incoming boundary layer developed. Each flow field of SWTBLI in the present experiment, using liquid nitrogen as the cryogenic coolant, was consistent with the classical Chaman's free interaction theory.

## 1. Introduction

Supersonic and hypersonic vehicles have recently been developed. Air-breathing engines, such as scramjet engines, are promising propulsion engines for high-speed vehicles. The supersonic intake or isolator, which is an important air-breathing engine component, decelerates and compresses the capture airflow through several oblique shock waves. Therefore, understanding shock wave/turbulent boundary layer interaction (SWTBLI) is important for the design of high-performance intakes or isolators, because SWTBLI may cause severe performance deterioration [1]. For example, the separation bubble in SWTBLI reduces the net cross-sectional area and results in unstart [1−3].

Moreover, the unsteadiness of SWTBLI is detrimental to maintaining flow stability [4]. Additionally, the structural and thermal loads are enhanced owing to separation [5]. Therefore, SWTBLI has been investigated and reviewed by several studies [2, 6−8].

The SWTBLI phenomenon is caused by shock waves with a strong adverse pressure gradient imposed on the turbulent boundary layer, whose momentum is lower. Consequently, the incoming turbulent boundary layer grows thicker and separates. The two main competing mechanisms of separation in SWTBLI were reported recently by Sabnis *et al.* [8], who reviewed the interaction physics of two-dimensional and three-dimensional flows. First, the pressure rise information from the shock wave can propagate upstream only through the subsonic region near the wall, which means that the size of the subsonic channel affects the upstream interaction length of SWTBLI and smears the pressure rise from inviscid flow. Therefore, while a higher-momentum boundary layer has more resistant to separation, a boundary layer with a smaller subsonic channel results in shorter interaction length, causing a stronger adverse pressure gradient and eventually resulting in less resistant to separation. Based on these two competing explanations of the separation mechanism, because the wall temperature leads to variations in the boundary layer profiles, such as velocity, density, and viscosity variations, the balance between the momentum and the subsonic channel is considered to be different to the adiabatic wall condition, leading to the modulation of the flow structure.

In actual supersonic intakes or isolators, the wall temperature $T_w$ does not always equal the recovery temperature $T_r$ (adiabatic wall temperature), meaning the wall-to-recovery temperature ratio $T_w/T_r \neq 1$. This is because the total temperature of the mainstream changes during actual flights. Some studies have suggested that, in scramjet engines with cryogenic fuel, the walls of supersonic intakes or isolators should be actively cooled using cryogenic fuel as the coolant to realize high-performance engine cycles [9, 10]. Therefore, it is necessary to understand the effect of the wall temperature on SWTBLI using theoretical, numerical, and experimental methods.

However, in the moderate supersonic flow ($M = 2 - 4$) considered in this study, the experimental data of the wall temperature effect on the SWTBLI, particularly visualized data, are insufficient for comparison with recent sophisticated numerical simulations [11−18], particularly regarding the heat transfer, pressure fluctuation, and flow structures of SWTBLI. The reason for this lies in the difficulty of creating a moderate wall cooling and heating in a supersonic wind tunnel from the perspective of experimental techniques. Some studies have reported flow visualization data under wall-heating conditions in moderate supersonic flow. Jaunet *et al.* [19] investigated the interaction length of SWTBLI at $M = 2.3$ with wall heating using particle image velocimetry (PIV) or schlieren visualization, and discussed Souverein's scaling law [20], which considers the wall temperature effect. Zhang *et al.* [21] visualized the turbulent boundary layer of heating and cooling, and visualized the SWTBLI at $M = 2.7$ with heating using the nano-tracer planar laser scattering technique and PIV. However, they did not obtain SWTBLI images under wall cooling. Zhou *et al.* [22] investigated the swept SWTBLI characteristics with wall heating using PIV at $M = 2.95$. Experimental studies on the wall-cooling effect in SWTBLI were initiated by Spaid *et al.* [23] at $M = 2.9$, or Back *et al.* [24] at $M = 3.5$ in the 1970s. However, to the best of the authors' knowledge, Hayashi *et al.* [25] have provided the only available flow visualization data on the wall-cooling effect at $M = 4$. They used the schlieren method and conducted heat transfer measurements under the same wall temperature conditions. In other words, they did not investigate the influence of wall temperature under the same mainstream conditions.

In previous hypersonic experimental studies at $M > 4$, which are often conducted in impulse wind tunnels, such as shock wind tunnels or expansion tubes, a wall-cooling state appears, and schlieren visualization is often conducted. This is because, in almost all cases, the total temperature of the free stream is very high, while the model surface temperature is typically equal to or slightly higher than room temperature. Consequently, the wall temperature $T_w$ is lower than the recovery temperature $T_r$, owing to the wall-cooling condition ($T_w < T_r$) [26, 27]. In previous hypersonic experimental studies other than impulse wind tunnels, only a few experiments in heated-air blowdown wind tunnel facilities [28, 29] were conducted. However, because there are differences, such as the real-gas effect [30] or turbulent boundary layer structures [31], between hypersonic flow and moderate supersonic flow, the dominant physical mechanisms in hypersonic flow on SWTBLI are different to those in supersonic flow.

Experimental data under wall-cooling conditions in moderate supersonic flow ($M = 2 - 4$), particularly visualized data, are still limited. Therefore, the objective of this study was to obtain limited experimental data on the wall-cooling effect in SWTBLI in supersonic flow, and understand the influence of wall temperature on actual vehicles. This paper presents test results including the schlieren visualization of the wall-cooling experiment using liquid nitrogen as a coolant in an air-breathing wind tunnel (designed Mach number $M = 2$) whose total temperature is room temperature. Furthermore, the flow field in these uncooling and cooling experiments is discussed based on Chapman's free interaction theory to confirm the validity of the experimental results obtained in this study.

## 2. Experimental set up and conditions

### 2.1. Supersonic wind tunnel

Figure 1 shows a supersonic wind tunnel with a rectangular cross-section. This wind tunnel is an atmospheric air-breathing type. Atmospheric air was inhaled into a vacuum tank through a silica gel dryer, two-dimensional nozzle, and test section. The vacuum tank volume was 50 m$^2$ and the tank pressure was set to approximately 5 kPa before each operation. The running time was maintained at > 8 s in this experiment. In the dryer, the pressure drop was less than 0.9 kPa. In the test section, which corresponds to the visualized section, the height was 38 mm, width was 80 mm, and length was 150 mm. The design Mach number in the test section was $M$ = 2.0. A wedge with a width of 79 mm and angle of 13° was set up to generate an oblique shock wave at the bottom wall of the test section. This oblique shock wave was imposed on the fully developed turbulent boundary layer on the upper wall.

The inner upper wall of the wind tunnel was flat and consisted of an aluminum alloy (A6063) with a thickness of 2.0 mm, which corresponded to the cooled region under the wall-cooling condition. Outside the upper wall, a liquid nitrogen pool was installed for wall cooling. Hence, in this experiment, the SWTBLI on the upper wall was investigated both with and without wall cooling. The detailed operation of the wall-cooling condition is explained in Section 3.2, along with the results of the thermocouple measurements. The windows in the test section were made of BK7 for schlieren visualization. The bottom and side walls were made of polycarbonate, because the thermal conductivity of this material is much lower than that of aluminum alloys. Thus, the windows and bottom and side walls can suppress cooling and prevent frosting during the precooling in the wall-cooling experiment.

Thermocouples were placed outside the upper wall surface, as presented in Fig. 1 and Table 1. The interior-wall surface temperature was estimated from the exterior wall temperature. Owing to the change in the wall temperature of the throat or side wall, the mainstream conditions may vary between the cooled and non-cooled cases. To evaluate the mainstream of the test section, the static pressure was measured at the location shown in Fig. 1 by replacing the BK7 window with an acrylic one that had static pressure holes.

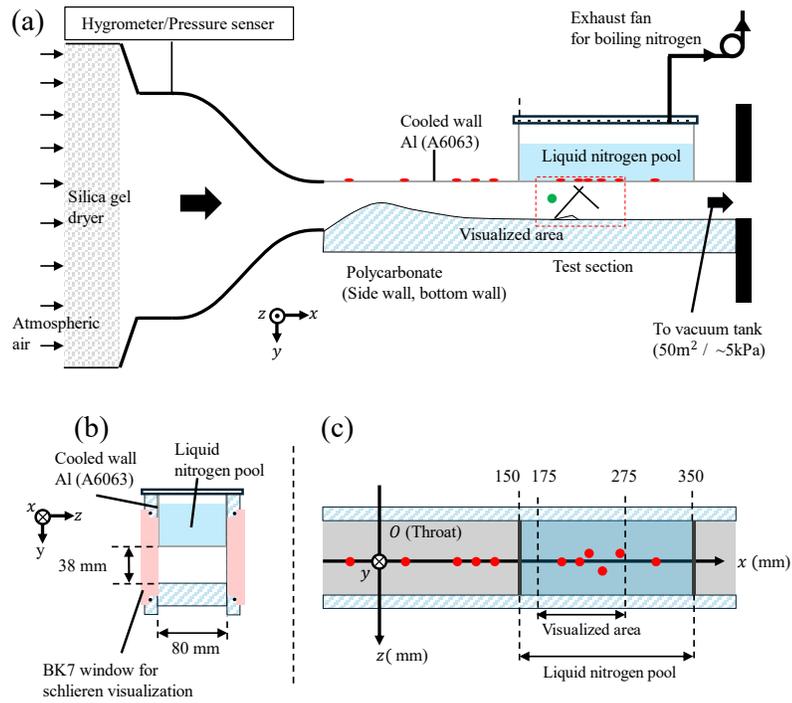

**Fig. 1.** Supersonic wind tunnel facility: (a) side view of overall schematic; (b) cross-sectional view of test section; (c) top view. Red points indicate thermocouple measurement points on upper wall's exterior surface. Coordinate origins were set as follows: $x$, at throat; $y$, at inner surface of upper wall; $z$, at center of span width. Green point at $(x, y) = (200 \text{ mm}, 20 \text{ mm})$ indicates static pressure measurement position in test section. Schlieren visualization was not conducted when measuring static pressure.

**Table 1.** Thermocouple measurement positions on upper wall's exterior surface; $x$ and $z$ axis defined as shown in Fig. 1; $T_{c1}$–$T_{c5}$ in upstream of liquid nitrogen pool; $T_{c6}$–$T_{c12}$ in liquid nitrogen pool.

|        | $T_{c1}$ | $T_{c2}$ | $T_{c3}$ | $T_{c4}$ | $T_{c5}$ | $T_{c6}$ | $T_{c7}$ | $T_{c8}$ | $T_{c9}$ | $T_{c10}$ | $T_{c12}$ |
|--------|----------|----------|----------|----------|----------|----------|----------|----------|----------|-----------|-----------|
| $x$(mm) | -35 | 25 | 80 | 100 | 120 | 190 | 210 | 220 | 235 | 255 | 295 |
| $z$(mm) | 0 | 0 | 0 | 0 | 0 | 0 | 0 | 5 | -5 | 5 | 0 |

### 2.2. Schlieren visualization system

The objective of schlieren visualization was to confirm the difference in the flow structure with and without wall cooling. Figure 2 shows the schlieren visualization system. The beam from the light source (Cavitar Ltd., CAVILUX Smart) was expanded at the pinhole and reflected from the convex mirror as a parallel beam. The beam passed over the test section and refocused on another convex mirror. A knife was installed at the focus point and the direction was set horizontally to clearly visualize the incoming boundary layer and reflected shock of SWTBLI. Consequently, the luminous intensity of the schlieren image became sensitive to vertical density gradient. A high-speed camera (Phantom Inc. V1211) captured the flow structure. In this experiment, because the operating time for each condition was longer than 8 s during visualization, the recording time was prioritized over the frame rate. Therefore, the frame rate was set to 100 fps, which was sufficient for capturing the effect of the wall temperature change on the flow structure of SWTBLI.

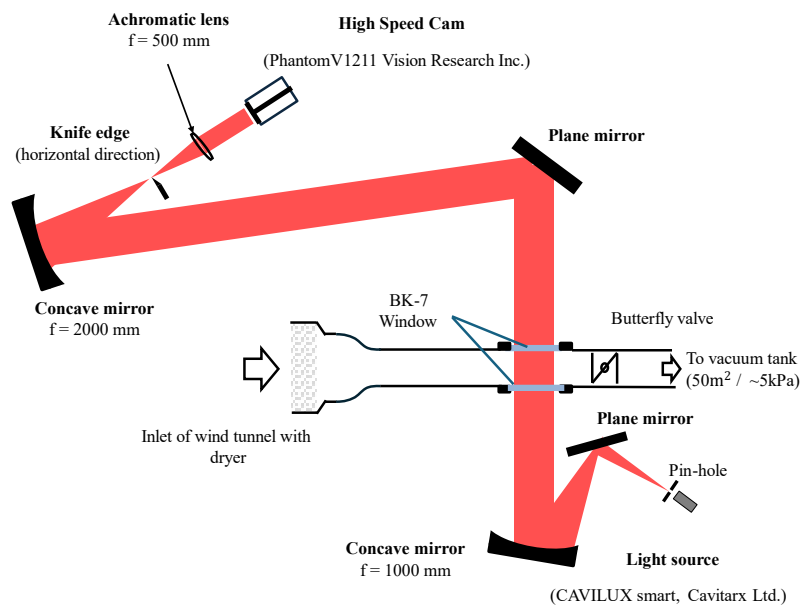

**Fig. 2.** Schlieren visualization system.

## 3. Results and discussion

In this section, the static pressure history is first described to confirm the mainstream conditions discussed in Section 3.1. In Section 3.2, the exterior wall surface temperature, interior upper wall surface temperature, and their time histories are discussed. The schlieren visualization results are reported, and the wall-cooling effect on SWTBLI observed in this experiment is discussed in Section 3.3.

### 3.1. Static pressure history

As discussed in Section 2.1, the mainstream may vary under cooling or non-cooling conditions. Figure 3(a) shows the static pressure history $P_{se}$ in the test section. The mean static pressures were $P_{se} = 12.3$ kPa without cooling and $P_{se} = 11.9$ kPa with cooling, respectively, and these standard deviations are within 0.1 kPa. The mainstream Mach number in the test section can be calculated from the static pressure by assuming an isentropic flow between the dryer and the test section. Figure 3(b) presents the mainstream Mach number histories, wherein the mean values in the operation were $M_e = 2.02$ without cooling and $M_e = 2.04$ with cooling. Therefore, because the difference between the cases with and without wall cooling is very small, it can be considered that the mainstream conditions are identical. Additionally, various other parameters in the test section, which were calculated using the same method requiring the Mach number from the static pressure, are listed in Table 2.

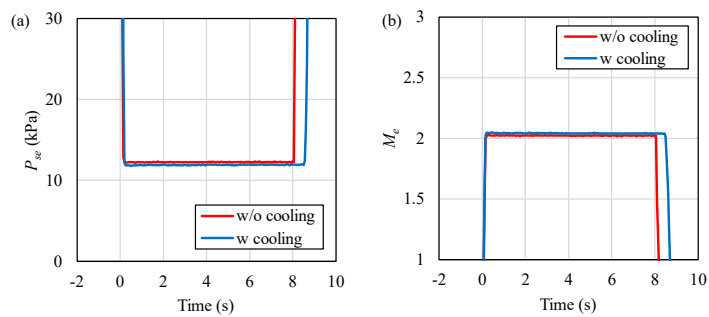

**Fig. 3.** (a) Static pressure history $P_{se}$ and (b) Mach number history $M_e$ without and with wall cooling in test section; operation begins at 0 s, and ends at 8 s or 8.2 s. Mach number $M_e$ is calculated under the assumption of isentropic flow between the dryer and the test section.

Table 2. Mainstream parameters in test section. These values were obtained when room temperature was 300 K, corresponding to total temperature in test section.

| Parameter | Symbol | Unit | Uncooling | Cooling |
| --- | --- | --- | --- | --- |
| Mach number | $M_e$ | - | 2.02 | 2.04 |
| Static pressure | $p_e$ | kPa | 12.3 | 11.9 |
| Static temperature | $T_e$ | K | 165 | 164 |
| density | $\rho_e$ | kg/m$^3$ | 0.260 | 0.254 |

### 3.2. Exterior and interior-wall surface temperature and cooling operation

In this section, the results of the exterior wall surface temperature $T_c$ measured by the thermocouples, particularly in the case of cooling, are discussed along with the operation. The inside-wall surface temperature $T_w$ is also discussed based on calculations from $T_{c6}$–$T_{c8}$ at the location where the fully developed boundary layer does not interact with the incident shock wave.

First, the calculation method for the interior-wall surface temperature $T_w$ and recovery temperature $T_r$ is described. Figure 4 shows the velocity and temperature distributions of the cooling condition, where a fully developed turbulent boundary layer exists. Considering that the heat transfer between the cooled wall and cooled boundary layer is one-dimensional, and that the turbulent boundary layer is fully developed, the inner surface temperature of the upper wall $T_w$ can be estimated as a function of the velocity profile parameter $n$ of the $1/n$-th-power law, as shown in Fig. 5. The step-by-step calculations are outlined in detail in the Appendix.

The recovery temperature $T_r$ is defined as follows:

$$T_r = T_e \left(1 + r\frac{\gamma - 1}{2} M_e^2\right) \quad (1)$$

where $T_e$ and $M_e$ are the static temperature and Mach number in the mainstream, respectively; $\gamma$ is the specific heat ratio and set to $\gamma = 1.4$ of the air. The recovery factor $r$, which depends on the Prandtl number, $P_r$, is set to $r = P_r^{1/3} = 0.896$ for the turbulent boundary layer. Therefore, the recovery temperature $T_r$ was 283 K because room temperature was 297 K when schlieren visualization and thermocouple measurement were conducted.

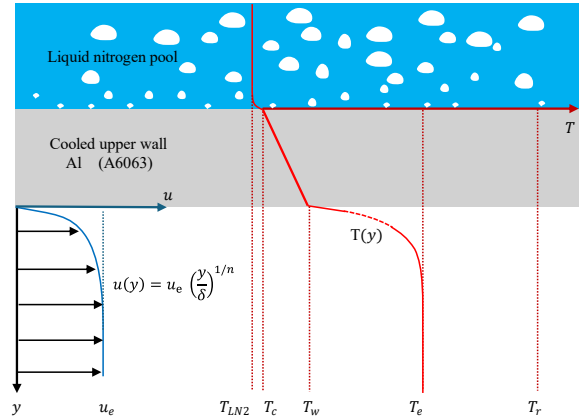

**Fig. 4.** Temperature and velocity distribution of cooled wall and cooled boundary layer in case of liquid nitrogen usage. The liquid nitrogen temperature $T_{LN2}$ was the saturation temperature of 77.35 K at atmospheric pressure. The wall surface temperature of the liquid nitrogen side $T_c$ was 78.5–80.5 K, as measured by the thermocouples.

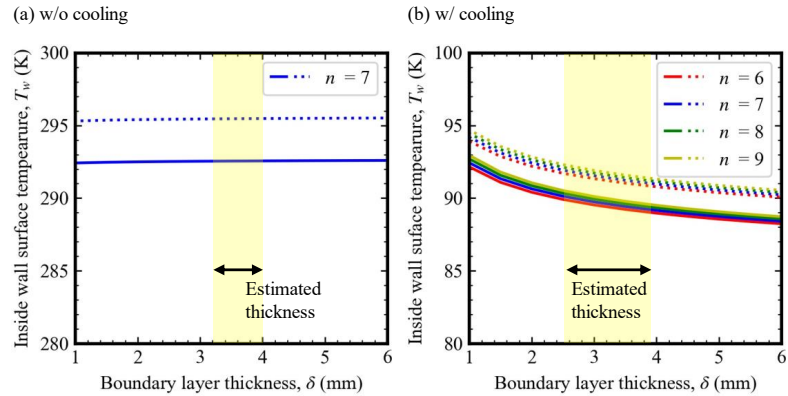

**Fig. 5.** The inner surface wall temperature of the upper cooled wall $T_w$ was calculated as described in the Appendix; $n$ is the $1/n$-th-power law of the velocity profile. (a) Case without cooling condition: solid line ($T_c = 296$ K); dotted line ($T_c = 293$ K). (b) Case with cooling: solid line ($T_c = 78.5$ K); dotted line ($T_c = 80.5$ K). The thickness was estimated from schlieren images.

Without cooling, the wind tunnel wall temperature was the same as the room temperature before operation. After the beginning of operation, the temperatures of all exterior walls ($T_c$) dropped by only 1–4 K from room temperature. Therefore, the interior-wall surface temperature $T_w$ was estimated at approximately 293–296 K, as shown in Fig. 5(a), where $T_c = 293, 296$ K and $n = 7$. Notably, Fig. 5(a) shows only the case of $n = 7$, because there is almost no difference in the range of $6 \leq n \leq 9$. The wall-to-recovery temperature ratio was $T_w/T_r \cong 1.04$–$1.05$, indicating that the non-cooled condition can be approximately considered as an adiabatic wall condition.

In this wall-cooling experiment, by pouring liquid nitrogen (77.35 K) into the pool on the upper wall, a wall-cooling state was created in the supersonic wind tunnel. The history of the upper wall surface temperature is shown in Fig. 6. The behavior of each thermocouple can be divided into two groups based on their locations: inside the liquid nitrogen pool ($T_{c6}$–$T_{c11}$) and outside the liquid nitrogen pool ($T_{c1}$–$T_{c5}$). At the beginning of operation, all temperatures matched the room temperature of 297 K. However, once the precooling began, the temperatures in the pool ($T_{c6}$–$T_{c11}$) decreased more rapidly compared with those outside the pool ($T_{c1}$–$T_{c5}$). Based on the rate of temperature drop in the pool, the poured liquid nitrogen exhibited film boiling from $-148$ s to $-75$ s. Subsequently, the boiling state transitioned to nuclear boiling. After $t = -50$ s, the pool temperature became steady at approximately 77.5 K, indicating that the heat transfer in the upper wall had reached a steady state. However, owing to the thermal conductive from the pool to the wind tunnel inlet, the temperature upstream of the pool ($T_{c1}$–$T_{c5}$) continued to decrease slightly. The running operation began at $t = 0$ s.

Immediately after the start of running operation, the temperature increased rapidly. Subsequently, during $t = 0$–$20$ s, while the external surface wall temperature in the pool ($T_{c6}$–$T_{c11}$) remained constant at approximately 78.5–80.5 K, which is the range of nuclear boiling, the temperatures upstream of the pool ($T_{c1}$–$T_{c5}$) gradually increased. Notably, the exterior temperature ($T_{c9}$–$T_{c11}$) at the location where the boundary layer is affected by the incident shock wave was almost the same as that in the upstream ($T_{c6}$–$T_{c8}$). At $t = 20$ s, the values in the pool increased rapidly because all liquid nitrogen boiled out, and the operation stopped at $t = 25$ s.

Figure 5(b) shows the calculated results for the interior-wall temperature $T_w$ below the liquid nitrogen pool with $T_c = 78.5, 80.5$ K and $n = 6, 7, 8, 9$. These $T_c$ values are the minimum and maximum values of $T_{c6}$–$T_{c8}$, respectively, during operation prior to the depletion of liquid nitrogen. Under the cooling condition, because the boundary layer thickness was approximately 2.5–3.9 mm, $T_w$ was approximately 88–92 K, corresponding to the wall-to-recovery temperature ratio $T_w/T_r = 0.31$–$0.33$. This estimated value is sufficient for investigating the wall-cooling effect on SWTBLI compared with those achieved in previous experimental studies, as shown in Fig. 7. The cryogenic characteristics of liquid nitrogen were efficiently used to create a sufficient wall-cooling state in supersonic flow of which total temperature was room temperature.

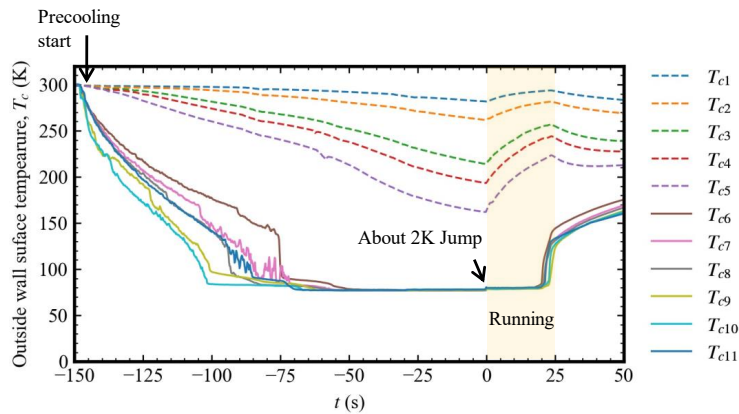

**Fig. 6.** Outside surface wall temperature time history measured by thermocouples with wall cooling. Dotted lines: upstream of liquid nitrogen pool ($T_{c1}$–$T_{c5}$); solid lines: inside liquid nitrogen pool ($T_{c6}$–$T_{c11}$).

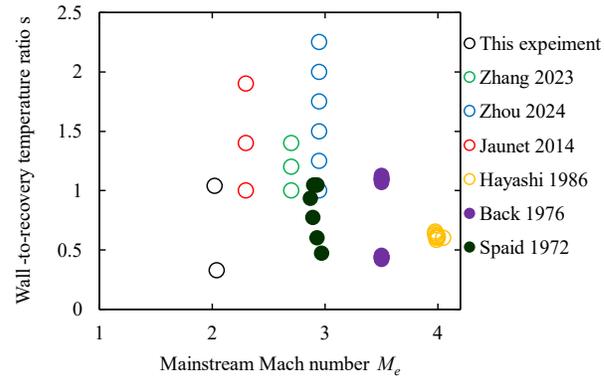

**Fig. 7.** Previous experimental conditions of SWTBLI in supersonic flow. Fill symbol: without visualization; Blank symbol: with visualization.

### 3.3. Schlieren visualization

This section presents the flow structure visualized through the schlieren method and discusses the impact of wall cooling on SWTBLI. Figure 8 shows a sketch of the typical SWTBLI in the schlieren images (Fig. 9) with and without cooling. Under both experimental conditions, the oblique shock wave generated by the wedge on the bottom wall interacted with the fully developed turbulent boundary layer on the upper wall. The thickness of the incoming boundary layer increased steeply from the foot of the separation shock wave. Moreover, the slip line appeared from the cross-point, suggesting that the incoming boundary layer should be separated. The reflected shock wave was generated from the crossing point and appeared slightly curved, owing to the expansion of the fan from the wedge. Therefore, the respective flow structures under the cooled and non-cooled conditions are typical SWTBLI with boundary separation [7].

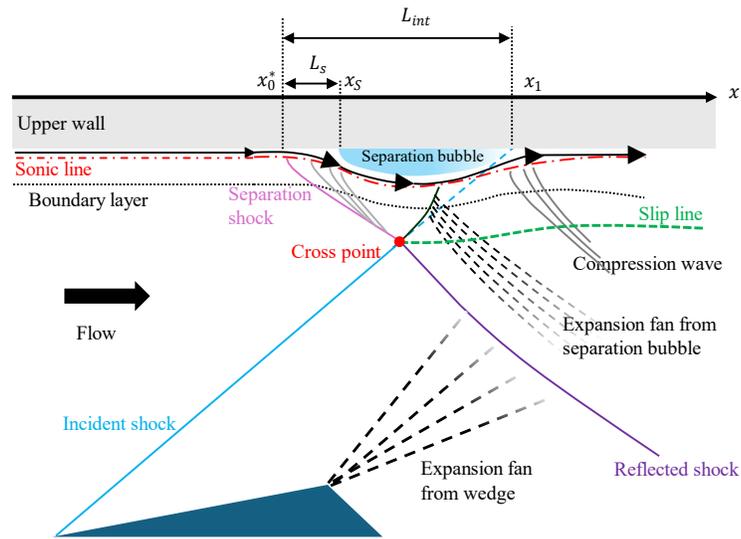

**Fig. 8.** Sketch of SWTBLI: $x_0^*$ is the separation shock foot, corresponding to the location of incipient pressure rise; $x_1$ is the location extrapolated from the incident shock; $x_s$ is the separation point.

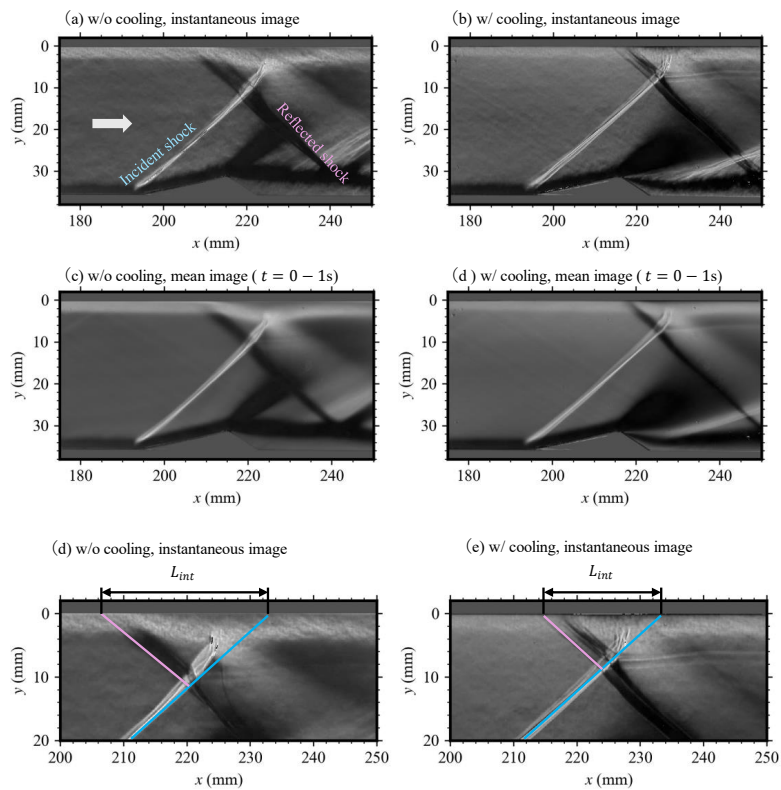

**Fig. 9.** Background processed schlieren images with horizontal knife edge direction. Instantaneous images: (a) without cooling and (b) with cooling, after $t = 0.04$ s from beginning of operation. Time-averaged images during $t = 0 - 1$ s from beginning of operation: (c) without cooling and (b) with cooling. Zoom instantaneous images of (a) and (b) with interaction length $L_{int}$: (d) without cooling and (b) with cooling.

However, noticeable differences exist in the reflected shock, separation shock, and incoming boundary layer thickness between the cooled and non-cooled cases.

Focusing on the reflected shock and separation shock, the shock positions with wall cooling moved downstream from the position without cooling, as shown in Fig. 9. Figure 10 shows a digital streak image that is useful for discussing the time history [32] of the reflected shock motion. This streak image was created by setting the pixel rows at $y = 14.6$ mm of each schlieren frame in chronological order. The white and black lines represent the trajectories of the incident and reflected shocks, respectively. While the white lines remain constant in both cases, the black lines shift by approximately 6–9 mm between the cooled and non-cooled conditions. Moreover, the inclination of the black line under the cooling condition indicates that the reflected shock wave gradually moved upstream.

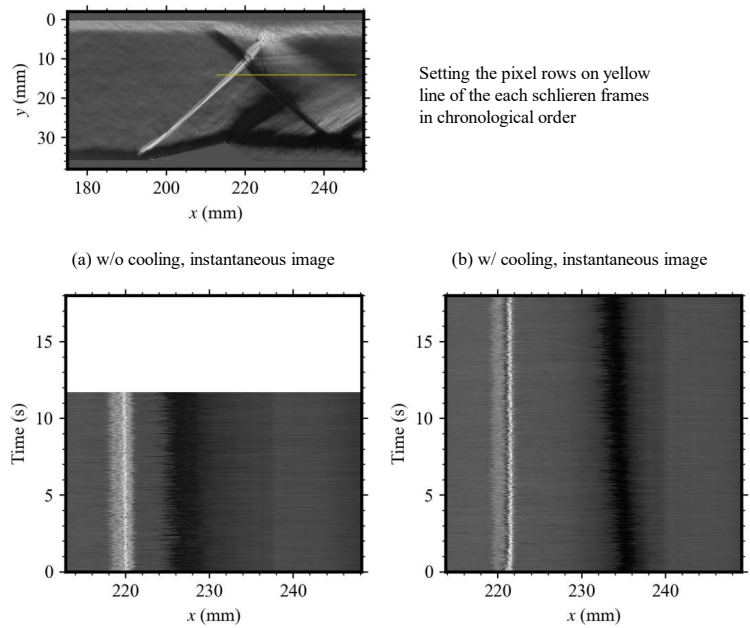

**Fig. 10.** Digital streak images from schlieren images: (a) without cooling and (b) with cooling. These streak images were created by setting the pixel rows on the yellow line of the schlieren frames in chronological order. The white line indicates incident shock; the black line indicates reflected shock.

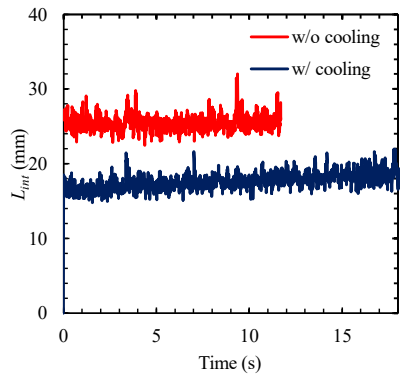

**Fig. 11.** Interaction length ($L_{int}$) histories of cooled and non-cooled cases. The separation shock position was determined to be the edge of the upstream side of the shock, whose luminous intensity was equal to 30 or more in the schlieren image.

Figure 11 shows the time histories of the interaction length $L_{int}$, which is defined as the length between the location where the incident shock is imposed on the wall in the inviscid flow and the location of the foot of the separation shock, similarly to Souverein *et al*. [20]. The schlieren images in Figs. 9(d) and (e) show $L_{int}$. As can be seen, the trends of the reflected shock motion are similar to those in Fig. 9. Specifically, under the cooling condition, $L_{int}$ was reduced by approximately 6–9 mm compared with that under the uncooled condition, and gradually increased. The $L_{int}$ suppression ratio of the cooling case to the uncooling case corresponds to approximately 0.60–0.72. Generally, the reflected shock and separation shock positions depend on the size of the separation or interaction length ($L_s$ or $L_{int}$), as shown in Fig. 8. In this experiment, based on the comparable results between the reflected shock and $L_{int}$ estimated from the separation shock, the separation under the wall-cooling condition was suppressed compared with the non-cooled condition. Moreover, despite gradual changing in the separation during the cooling case, the wall surface temperature near the separation $T_w$ remained constant. This steadiness makes the results for the reflected shock and $L_{int}$ from the separation shock in the wall-cooling case particularly noteworthy.

By comparing the incoming boundary layer on the upstream side of the separation with and without cooling in the schlieren images (Fig. 9(d) and (e)), it can be found that the boundary layer of the cooling case is smaller than that of the uncooled case. The thickness of the cooled and non-cooled case was approximately 3.2–4.0 mm and approximately 2.5–3.9 mm, respectively, as evaluated from Fig. 9(d, e). The reason for the difference between each case can be inferred from the growth of the boundary layer. Figure 12 shows the time history of the incoming boundary layer thickness in the cooling case. The boundary layer thickness of Fig. 12 is defined as the thickness from the wall to the edge of the region where the luminance value is 130 or more. As observed in Fig. 12, the thickness of the incoming boundary layer gradually increased. Notably, the surface temperature of the interior wall near the separation region $T_w$ remained almost unchanged. This finding is interesting, particularly in conjunction with the results related to the motion of the reflected shock and interaction length $L_{int}$. This phenomena can be guessed to be due to the gradual increase of the upstream wall temperature from the liquid nitrogen pool ($T_{c1} - T_{c5}$), where the boundary layer developed, as shown in Fig. 6. Consequently, the reflected shock and interaction length $L_{int}$ may have also modulated during operation. Therefore, these results suggest the precooling and the running time affect the incoming boundary layer.

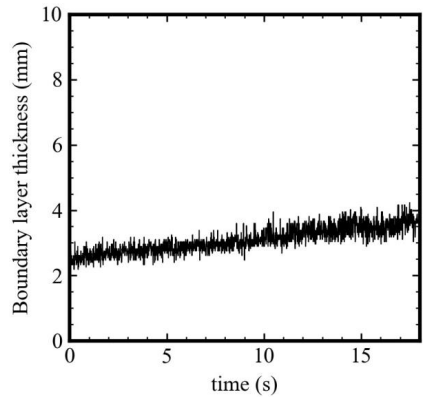

**Fig. 12.** Time history of incoming boundary layer at $x = 200$ mm in schlieren image under cooling condition.

The phenomena of reflected shock and separation shock motion can be explained by Chapman's classic free interaction theory [33]. This theory explains that an interaction length $L_s$ from the separation point (Fig. 8) depends on the Mach number $M_e$, boundary layer displacement thickness $\delta_0^*$, and upstream skin friction coefficient $C_{f0}$, as follows:

$$L_s \propto \frac{\delta_0^*}{\sqrt{C_{f0}}} \frac{1}{\sqrt[4]{M_e^2 - 1}}. \tag{2}$$

The equation suggests that, if the mainstream is constant, the interaction length $L_s$ depends only on the incoming boundary layer parameters $\delta_0^*$ and $C_{f0}$. Similar to previous results, under the wall-cooling condition, the friction coefficient increased, while the boundary layer displacement thickness decreased. This implies that $L_s$ decreased owing to the wall-cooling effect. Therefore, by assuming that the variation of the reflected shock position and interaction length $L_{int}$ scaled with $L_s$ [7], it can be inferred that changes in the incoming parameters $\delta_0^*$ and $C_{f0}$ influenced the reflected shock motion and interaction length $L_{int}$. The hypothesis that the variation of the reflected shock and interaction length $L_{int}$ during the cooling experiment responded to fluctuations in the incoming boundary layer is consistent with Chapman's free interaction theory.

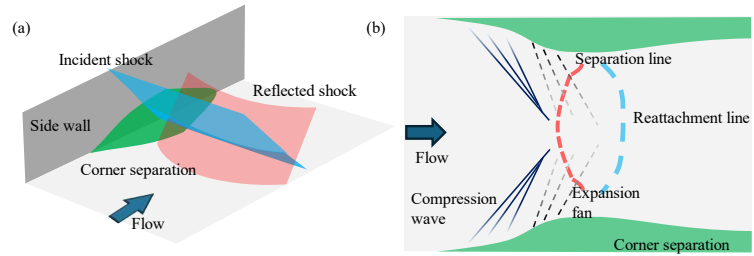

**Fig. 13.** Schematic diagrams of corner effect with distorted reflected shock along spanwise direction: (a) overview; (b) top view. The reflected shock and separation lines are curved owing to the compression wave and expansion fan from the corner separations.

This explanation, based on Chapman's classic free interaction theory, holds true exclusively for pure two-dimensional flow. However, owing to the existence of a side wall, the real flow structure of SWTBLI in this experiment exhibited three-dimensional characteristics, similar to corner effects [34, 35]. If separation occurs in the wind tunnel's corner, as shown in Fig. 13, compression waves and expansion fans are generated from the corner separation and influence the SWTBLI at the center of the wind tunnel. Consequently, the separation line or reflected shock is distorted along the spanwise direction. Additionally, the size of the separation at the center with the corner effect is occasionally larger than that in the case of pure two-dimensional flow. In this experiment, this corner effect can be estimated qualitatively from the thickness of the reflected shock in the schlieren image (Fig. 9), because the shock thickness depends on distortion along the spanwise direction [33−35]. The thickness of the reflected shock in the wall-cooling case is clearly different to that in the case without cooling (Figs. 9 and 10), suggesting that the corner effect was mitigated more effectively under wall cooling. Corner separation suppression is assumed to have occurred because the boundary layer on the corner also decreased owing to the cooling of the upper wall. Therefore, the suppression of boundary layer separation on the upper wall may have resulted not only from the cooling effect, which is typically considered as a pure two-dimensional flow, but also from the reduction of corner effects. In other words, this approach could help mitigate performance degradation in real intakes or isolators caused by complex three-dimensional separation flows.

## 4. Conclusion

This paper presents the testing results obtained by a wall-cooling experiment using liquid nitrogen in an air-breathing wind tunnel of which the total temperature was maintained at room temperature. By leveraging the cryogenic characteristics of liquid nitrogen, sufficient wall cooling for several seconds under supersonic flow conditions ($M_e = 2.02\text{--}2.04$) was achieved. The findings of this study are summarized as follows:

In the case of wall cooling, the temperatures of the exterior surface of the cooled wall were divided into two groups based on the measurement positions inside and outside the liquid nitrogen pool. The temperature of the former position was almost steady and approximately 78.5–80.5 K during the wind tunnel's operation. However, the latter position exhibited an increase in drift when the wind tunnel operated. The interior temperature during operation was calculated as 293–296 K in the case without cooling and 88–92 K in the case with cooling, corresponding to the wall-to-recovery temperature ratios ($T_w/T_r$) of 1.04–1.05 and 0.31–0.33, respectively.

With wall cooling, the reflected shock position moved downstream in relation to the position without wall cooling. The shock trajectory with wall cooling, as observed in the digital streak image from the schlieren image, gradually moved upstream. The interaction length $L_{int}$ determined by the separation shock foot and the point where the incident shock impacted the wall was obtained alongside the reflected shock motion. Specifically, in the wall-cooling case, $L_{int}$ was smaller than that in the case without cooling and gradually increased. The results for the reflected shock motion and $L_{int}$ indicate that the separation was suppressed, and the ratio of the cooling case to the uncooling case is approximately 0.60–0.72. Therefore, this experiment utilizing liquid nitrogen effectively demonstrated the suppression of the separation flow field resulting from the wall-cooling effect.

The thickness of the incoming boundary layer under wall cooling was smaller than that without cooling. These thicknesses were approximately 3.2–4.0 mm and approximately 2.5–3.9 mm, respectively. Hence, it can be inferred that the upstream wall cooled by the liquid nitrogen pool suppressed the growth of the boundary layer. Additionally, during the operation under the cooling condition, the thickness of the incoming boundary layer gradually increased, which is attributed to the rising temperature of the upstream upper wall. Therefore, the fluctuation in the growth of the incoming boundary layer, resulting from changes in the wall temperature distribution, influenced the flow structure of SWTBLI, including the reflected shock and interaction length $L_{int}$ related to separation shock, as observed in this experiment. This understanding of the response of the flow structure to wall cooling is consistent with the classical Chapman's free interaction theory, which is valid for pure two-dimensional flow.

From the viewpoint of the three-dimensional flow, the flow was somewhat affected by the sidewall based on the reflected shock thickness in the schlieren image. A qualitative comparison between the cases with and without cooling revealed that corner separation with wall cooling may have been suppressed. This suggests that wall cooling near the corner probably contributed to the effective suppression of boundary layer separation, owing to the change in the structure of three-dimensional flow.

The findings of this study confirm that using liquid nitrogen as a coolant is valid for investigating the wall-cooling effect on SWTBLI. Conducting further measurements of the incoming boundary layer or SWTBLI flow field, including the three-dimensional characteristics, can yield more accurate experimental data and a deeper understanding of the wall-cooling effect.

## Acknowledgement

This work was supported by JSPS KAKENHI Grant Numbers 20H02350 and by JST SPRING, Grant Number JPMJSP2125. Yuma Miki, would like to take this opportunity to thank the "THERS Make New Standards Program for the Next Generation Researchers." The authors also would like to appreciate Professor Akihiro Sasoh for his valuable advice.

## Author Declarations

### Conflict of Interest

The authors have no conflicts to disclose.

### Data Availability

The data that support the findings of this study are available from the corresponding author upon reasonable request.

**Appendix: estimation of interior-wall surface temperature**

Considering that the heat transfer in the cooled wall is one-dimensional, as shown in Fig. 4, the interior surface temperature can be estimated. In this experiment, the known temperature values for the heat transfer were the exterior wall temperature $T_c$ values measured by thermocouples, and the recovery temperature $T_r$. The $T_r$ value was calculated from the main flow values as expressed by Eq. (1). The equations for the heat conduction of the cooled wall and heat convection of the compressible boundary layer are as follows:

$$q = \frac{T_w - T_c}{R_w} \quad (A1)$$

$$q = h_{BL}(T_w - T_r) \quad (A2)$$

where $q$ is the heat flux, $R_w$ is the thermal resistance, and $h_{BL}$ is the heat transfer coefficient. Because the required value is $T_w$, the equation to be solved is derived from Eqs. (A1) and (A2).

$$T_w = \frac{\frac{1}{R_w}T_c - h_{BL}T_c}{\frac{1}{R_w} - h_{BL}}. \quad (A3)$$

The values of thermal resistances $R_w$ and $T_c$ are known. Therefore, if $h_{BL}$ is required, $T_w$ can be obtained as follows:

First, because the only known information about the boundary layer profile is thickness $\delta$ from the schlieren images and main flow velocity $u_e$, the boundary layer velocity profile is assumed to obey the $1/n$ power law, as follows:

$$\left(\frac{u}{u_e}\right) = \left(\frac{y}{\delta}\right)^{\frac{1}{n}}. \quad (A4)$$

Furthermore, it is assumed that the Walz equation, which gives the temperature–velocity relationship, holds.

$$\frac{T}{T_e} = \frac{T_w}{T_e} + \frac{T_r - T_w}{T_e}\left(\frac{u}{u_e}\right) - r\frac{(\gamma-1)}{2}M_e^2\left(\frac{u}{u_e}\right)^2. \quad (A5)$$

Momentum thickness $\theta$ is defined as follows:

$$\frac{\theta}{\delta} = \int_0^1 \frac{\rho u}{\rho_e u_e}\left(1 - \frac{u}{u_e}\right) d\left(\frac{y}{\delta}\right). \quad (A6)$$

Therefore, according to Eqs. (A4)–(A6), the momentum thickness $\theta$ can be calculated as a function of $T_w$ and $n$. In other words, the Reynolds number based on the momentum thickness $Re_\theta = \rho_e u_e \theta / \mu_e$ can be expressed as variables of $T_w$ and $n$. A similar calculation under the assumption of the $1/7$ power law and a given temperature–velocity relationship has been reported by Bartz *et al.* [39].

Next, skin friction prediction theories are used to transform a compressible–incompressible state to use semi-empirical laws, such as the Karman–Schoenherr equation:

$$\frac{1}{C_{f_{inc}}} = 17.08 \left( \log_{10} Re_{\theta_{inc}} \right)^2 + 25.11 \log_{10} Re_{\theta_{inc}} + 6.012. \tag{A7}$$

Skin friction coefficient $C_f = 2\tau_w / \rho_e u_e^2$ can be found from the known Reynolds number $Re_\theta$. Hopkins et al. [40] reviewed various prediction theories that recommend using the Van Driest II transformation. In this approach, the transformed incompressible skin friction coefficient, $C_{f_{inc}}$, and the incompressible Reynolds number based on the momentum thickness $Re_{\theta_{inc}}$ are defined as $C_{f_{inc}} = F_c C_f$ and $Re_{\theta_{inc}} = F_\theta Re_\theta$, respectively. Transformation functions $F_c$ and $F_\theta$ are expressed as follows:

$$F_c = \frac{m}{(\sin^{-1}\alpha + \sin^{-1}\beta)^2} \tag{A8}$$

$$F_\theta = \frac{\mu_e}{\mu_w} \tag{A9}$$

where

$$m = 0.2 r M_e^2 \tag{A10}$$

$$\alpha = \frac{2A^2 - B}{\sqrt{4A^2 + B^2}} \tag{A11}$$

$$\beta = \frac{B}{\sqrt{4A^2 + B^2}} \tag{A12}$$

$$A = \sqrt{\frac{m}{\frac{T_w}{T_e}}} \tag{A13}$$

$$B = \frac{1+m}{\frac{T_w}{T_e}} - 1 \tag{A14}$$

where $\mu_e$ and $\mu_w$ are the viscosity in the free stream and vicinity of the wall. In this estimation, these viscosities follow the Sutherland formula, as follows:

$$\frac{\mu}{\mu_0} = \left(\frac{T}{T_0}\right)^{\frac{3}{2}} \frac{T_0 + S}{T + S} \tag{A15}$$

where $\mu_0 = 1.7894 \times 10^{-5}$ Pa s, $T_0 = 288.15$ K, and $S = 110.4$ K. Notably, transformation functions $F_c$ and $F_\theta$ depend on $T_w$, as expressed by Eqs. (A8)–(A15). In summary, $Re_{\theta_{inc}} = F_\theta Re_\theta$ are variables of $T_w$ and $n$. Moreover, from the Karman–Schoenherr equation expressed in Eq. (A7), $C_f$ can also be expressed as variables of $T_w$ and $n$. Consequently, $C_f$ can be treated as a function of $T_w$ and $n$ if $Re_\theta$ is a known function depending on $T_w$ and $n$.

Finally, this study used the Reynolds analogy relationship to obtain $h_{BL}$ from $C_f$. The Reynolds analogy is expressed as follows:

$$C_h = \frac{s}{2} C_f \qquad (A16)$$

where Stanton number $C_h = h_{BL}/\rho_e c_p u_e$ and analogy factor $s = P_r^{-2/3} = 1.24$. Therefore, $h_{BL}$ can be considered as a function of $T_w$ and $n$, meaning that $T_w$ can be determined by solving Eq. (A3) with respect to each velocity profile parameter $n$.